\documentclass[11pt]{article}
\usepackage[a4paper,hmarginratio=1:1,vmarginratio=2:3,
totalwidth=15.6cm,totalheight=21.6cm]{geometry}
\usepackage{bm,epstopdf,epsfig,amsmath,amssymb,amsfonts,colordvi,latexsym,comment,cancel,
verbatim}
\usepackage{graphicx,graphics}

\usepackage[usenames,dvipsnames]{color}
 
\usepackage{setspace}

\newcommand{\cD}{{\cal D}}

\newcommand{\cJ}{{\cal J}}

\newcommand{\cL}{{\cal L}}
\newcommand{\cM}{{\cal M}}

\newcommand{\cO}{{\cal O}}

\newcommand{\tr}{\text{tr}}

\newcommand{\ra}{\rightarrow}
\newcommand{\be}{\begin{equation}}
\newcommand{\ee}{\end{equation}}
\newcommand{\bea}{\begin{eqnarray}}
\newcommand{\eea}{\end{eqnarray}}

\newcommand{\opsi}{\overline\psi}
\newcommand{\baa}{\begin{array}}
\newcommand{\eaa}{\end{array}}
\long\def\symbolfootnote[#1]#2{\begingroup
\def\thefootnote{\fnsymbol{footnote}}\footnote[#1]{#2}\endgroup}
\begin{document}

\thispagestyle{empty}
\begin{flushright}
CERN-PH-TH/2015-292\\
\today
\end{flushright}

\vspace{2.cm}

\begin{center}
{\Large {\bf Comments on the nilpotent  constraint

\bigskip
  of the goldstino superfield}} 
 \medskip
 \vspace{1.cm}

D. M. Ghilencea$^{\,a,\,b}$
\bigskip

$^a$ {\small CERN Theory Division, CH-1211 Geneva 23, Switzerland} 

$^b$ {\small Theoretical Physics Department, National Institute of Physics and}

{\small  Nuclear Engineering (IFIN-HH) Bucharest, MG-6 077125,  Romania}

\end{center}

\bigskip

\begin{abstract}
\noindent
Superfield constraints were often  used in the past, in particular to  describe the 
Akulov-Volkov action of the goldstino by a superfield formulation with
$L\!=\!(\Phi^\dagger \Phi)_D\!+ [(f\Phi)_F\!+\!{\rm h.c.}]$ endowed with  
the  nilpotent constraint $\Phi^2=0$ for the goldstino superfield ($\Phi$). 
Inspired by this, such constraint is often used to define
the goldstino superfield  even {\it in the presence} of  additional superfields, for example
 in models of ``nilpotent inflation''.
In this review we show that  the  nilpotent property  
is not valid in  general, under the  assumption of a  microscopic  (ultraviolet)  
description of  the theory with  linear supermultiplets.
Sometimes only weaker versions  of the nilpotent relation are true 
such as $\Phi^3=0$ or $\Phi^4=0$  ($\Phi^2\not=0$)  in the {\it infrared}
(far below the UV scale) under the further requirement of decoupling all
additional scalars (coupling  to  sgoldstino),  something not always possible
(e.g. if light scalars exist). In such cases the weaker nilpotent property is not
specific to the goldstino superfield anymore.
We review  the restrictions for the Kahler curvature tensor and superpotential $W$
under which  $\Phi^2=0$ remains true in infrared,  assuming linear 
supermultiplets in the microscopic description. One can reverse the 
arguments to  demand that the nilpotent condition, initially an infrared property, 
be extended even in the presence of  additional superfields, but
this may question the nature of supersymmetry breaking or the existence 
of a perturbative  ultraviolet  description  with linear supermultiplets.
\end{abstract}

\newpage

\section{Introduction}

Superfield constraints were often used in the past (see \cite{Rocek} for early models) 
on microscopic (ultraviolet) supersymmetric Lagrangians  to project out some of the 
degrees of freedom (of that superfield)
and  to obtain in this way non-linear realizations of supersymmetry.
 For the  case of a single superfield, 
an interaction-free Lagrangian 
$L=\int d^4\theta \Phi^\dagger \Phi+(\int d^2\theta \, f\Phi+h.c)$
 endowed with the nilpotent  constraint $\Phi^2=0$ provides \cite{Gatto}
a simple  superfield description of  the famous Akulov-Volkov Lagrangian \cite{AV}, 
see also more recent \cite{SK}.
Here $\Phi$ is the goldstino superfield, $\Phi=(\phi,\psi,F)$ where $\phi$ is the 
 sgoldstino, the scalar superpartner of goldstino $\psi$. The solution to the nilpotent
 constraint is $\Phi=\psi\psi/(2 F)+\sqrt 2 \theta\psi
+\theta\theta F$ which when
used in $L$ recovers {\it onshell} the Akulov-Volkov action \cite{Kuzenko,Kuzenko2}.
Actually $L$  was itself derived by starting from a  general 
Kahler potential  $K(\Phi,\Phi^\dagger)$ and superpotential $W(\Phi)$ 
after  taking the limit of an {\it infinite} sgoldstino mass \cite{Gatto} giving 
$\phi=\psi\psi/(2F)$  and thus leading to  $L$.
Further, the Akulov-Volkov result was extended to supergravity \cite{ED1,ED2}.

The nilpotent property $\Phi^2=0$  was then used  beyond  the Akulov-Volkov action, 
even in cases when additional  superfields are present \cite{SK}. 
Its applications to supersymmetric models were studied together with other 
projector relations applied to the microscopic   Lagrangian, to decouple either
bosonic or fermionic  superpartners, that we do not discuss here. 
More interestingly, it was noticed in \cite{SK}  and verified in general models
 \cite{dg1}  that the goldstino superfield  is the infrared limit
of the  superconformal symmetry breaking  chiral superfield $X$ \cite{CL,JD}
that breaks the conservation of the Ferrara-Zumino  supercurrent \cite{FZ}.

In this review, we start  from  a general microscopic 
(UV) Lagrangian with linear supersymmetry {\it in the presence} of additional 
superfields $\Phi_i=(\phi_i,\psi_i,F_i)$ and investigate if one can actually have
$\Phi^2=O(1/\Lambda)$ for the goldstino superfield, with $\Lambda$ the UV scale
(related to the Kahler curvature tensor). 
This would give $\Phi^2=0$ {\it in the infrared} i.e.  at vanishing momenta and
scales far below  $\Lambda$.
We show that the  answer strongly depends on the properties of 
the Lagrangian. The  nilpotent  property of the
 goldstino superfield means that the sgoldstino is decoupled
(i.e. it is massive enough to be integrated out via equations of motion).
In general  this is not always possible since
the sgoldstino is often light or  even massless at tree 
level\footnote{as in the O'Raifeartaigh model.}  or perturbation theory breaks down.
Moreover  the sgoldstino is often a mixture of many scalar fields. Only
 upon the integration of all these scalars (if massive enough) could one hope for  a solution of
 the form   $\phi=\psi\psi/(2 F)$ and thus for $\Phi^2$ to  vanish in the infrared.
However,  in  general $\phi=a_{ij}\,\psi^i\psi^j+a^{kl}\opsi_k\opsi_l
+c_{ij}^{kl}\psi^i\psi^j\,\opsi_k\opsi_l$
 giving
$\Phi^2\not=0$, unless additional conditions are met.
These issues are often ignored  in the literature. The purpose of this work  is to 
review   the  additional restrictions to be respected  by the Kahler potential $K$ and 
superpotential $W$ in order to have $\Phi^2=0$ in the infrared.

One can also reverse the arguments and take a different, 
 easier approach: simply take the nilpotent constraint
as  a {\it definition} for the goldstino superfield even in the presence  of additional 
superfields,   without being concerned  about the 
existence of a UV  linear  description of the goldstino multiplet.  One thus gives 
up the UV  microscopic description, which might not even exist (if  restrictions like those
 discussed earlier are not respected).
There are various applications of this approach, e.g. \cite{L1,dg4,F1,K1,D1,D2,LQ,FK,HY}.
This method is particularly popular in some models of
 ``nilpotent inflation''  
because in this case one does not need to stabilize the sgoldstino direction
since it is a bilinear of fermions, leading to a  simplification
of  problems like  moduli stabilization.
Note however that the nilpotency of goldstino was initially an infrared property \cite{SK,dg1}, 
valid at vanishing momenta and scales far below $\Lambda$ (or the scale
of inflation, etc). Questions also  remain on the exact  nature of supersymmetry  breaking 
and on the existence of a perturbative  ultraviolet completion with linear supermultiplets.

Here  we adopt the view of a microscopic Lagrangian with linear supermultiplets as the starting point.
We review simple counter-examples to the condition $\Phi^2=0$ in infrared 
that show that even in the minimal case of two superfields
such constraint is not respected and cannot be used to define the goldstino superfield.
For simple  $K$ and $W$
  weaker versions of this property are often true, such as $\Phi^3=0$ or $\Phi^4=0$ with
$\Phi^2\not=0$ and only  {\it after decoupling both scalars} of the theory.
Moreover this weaker property applies to both superfields i.e. it is not specific to the 
goldstino superfield. The reason for the weaker nilpotent relation
 is simple: for large enough powers, $\Phi^n$ vanishes 
due to the presence of a large power of Grassmann variables; this happens if
 both scalars of the two superfields are massive enough (to be integrated out) 
to become combinations of  Weyl bilinears ($\phi=a_{ij}\psi^i\psi^j
+a^{kl}\opsi_k\opsi_l+c_{ij}^{kl}\psi^i\psi^j\opsi_k\opsi_l$).
 The case of a light  (matter) scalar
in the model can  invalidate even these weaker constraints.
Assuming linear  supermultiplets in a microscopic  $K$ and $W$,
the infrared nilpotent property of the goldstino superfield is maintained under
restrictive conditions for the Kahler curvature tensor and $W$ that we shall identify.
Otherwise  $\Phi^2\not=0$  in the infrared.

For general $K$ and $W$, identifying the goldstino superfield can be difficult.
One has to identify the ground  state and  the sgoldstino can be a complicated 
function of the other scalars. This issue  can be avoided in applications since the 
superconformal symmetry breaking chiral
 superfield $X$ goes in the infrared to the goldstino superfield
 $X \ra (8 f/3)\,\Phi$, see \cite{SK,dg1}. Here $\sqrt f$ is the 
supersymmetry breaking scale  and $X$ is the solution to the equation
$\overline D^{\dot \alpha} \cJ_{\alpha\dot\alpha}=D_\alpha X$ with
$X=(\phi_X,\psi_X,F_X)$;  $J$ is the Ferrara-Zumino multiplet of currents \cite{FZ}.
Further, for given $K$ and $W$ one has \cite{CL,JD}
\bea\label{X}
X=4 W(\Phi^j) -\frac{1}{3}\,{\overline D}^2 K(\Phi^j,\Phi^\dagger_j)
-\frac{1}{2}\,{\overline D}^2 Y^\dagger(\Phi_j^\dagger)
\eea
where $\overline D^2 Y^\dagger$ is an improvement term. 
This expression  can be used in applications to identify the sgoldstino in the infrared
and  also  to see if  $\Phi^2=\cO(1/\Lambda)$ after integrating the scalars.
For the examples considered  we  verify that in the infrared 
$X$ goes to the goldstino superfield.

\section{Nilpotent goldstino superfield: some (counter)examples}

\subsection{A simple model}
\label{ex1}

Let us first review a simple  model \cite{SK}
\medskip
\bea\label{st1}
K=\Phi^\dagger\Phi-\frac{c}{\Lambda^2}\,\Phi^2\,\Phi^{\dagger 2}
-\frac{\tilde c}{\Lambda^2}\,(\Phi^3\,\Phi^\dagger +\Phi\,\Phi^{\dagger 3}\,)
+\cO(1/\Lambda^3),
\quad W=f\,\Phi
\eea

\medskip\noindent
Supersymmetry is broken by non-zero  $\langle F\rangle$, so
 $\Phi=(\phi,\psi,F)$ is a  goldstino superfield.
The higher dimensional D-terms provide a mass term for the sgoldstino $\phi$, 
while the goldstino is massless, as expected.
The scalar potential is
\medskip
\bea
V= f^2\,\big[
1+4 \,c/\Lambda^2\,\,\phi^\dagger\phi+3 \tilde c/\Lambda^2\,\,( \phi^2+{\rm h.c.})
+\cO(1/\Lambda^3)
\big]
\eea

\medskip\noindent
The  masses of the two real scalars $\varphi_{1,2}$ of $\phi$ 
are: 
$m_{1,2}^2=f^2\,(4\,c\pm 6 \tilde c)/\Lambda^2$. Thus one must choose  
$\vert\tilde c\vert < (2/3)\,c$ to ensure positive scalar (masses)$^2$.
Since sgoldstino is massive and goldstino is massless, one can integrate out
the former by using its equation of motion at zero momentum and then  expanding it 
about the ground state, to find
 \medskip
\bea\label{rr2}
\phi=-\frac{\psi\psi}{2\,f}+\cO(1/\Lambda)
\eea

 \medskip\noindent
where   $\cO(1/\Lambda)$ denotes  terms suppressed by $\Lambda$, 
e.g.  $\sqrt f/\Lambda$, etc.
So the sgoldstino  as a composite of  goldstini.
Note that the limit of restoring supersymmetry
$f\ra 0$ in eq.(\ref{rr2}) does not exist.
Further, with  $\phi$ as in eq.(\ref{rr2}), one has that
 \bea
\Phi^2=\cO(1/\Lambda)
\eea
i.e. its square is vanishing in the infrared limit
(defined here by zero momentum and $\Lambda$ much larger than all 
other scales in the theory). 
This limit should be taken with care 
since it  is not always well defined  perturbatively. 
Indeed, one has $m_{1,2}^2\sim f$ (from supersymmetry breaking) which together with
$m_{1,2}^2 < \Lambda^2$ give $\Lambda^2/\rho\sim f < \Lambda^2/\sqrt \rho$,
for $\rho=4c\pm 6\tilde c>0$. This range for $f$ is very small  
and if $\cO(\sqrt f/\Lambda)$ cannot be neglected then $\Phi^2$ is not vanishing.

From eq.(\ref{X}) after using the equation of motion $\overline D^2\Phi^\dagger
=4\,f+\cO(1/\Lambda^2)$, one has (ignoring the improvement term)
\bea
X=4 W-\frac{4}{3} f\,\Phi+\cO(1/\Lambda^2)=\frac{8}{3}\,f\,\Phi+\cO(1/\Lambda)
\eea
This verifies that in the infrared and onshell $X\ra (8/3)\,f\, \Phi$.

\subsection{O'Raifeartaigh model and nilpotent goldstino superfield}
\label{ex2}

Let us now consider a model  that includes other fields in addition
to the goldstino superfield, such as the O'Raifeartaigh model  
\medskip
\bea\label{ae1}
K = \Phi_1^\dagger\Phi_1+ \Phi_2^\dagger\Phi_2+\Phi_3^\dagger\Phi_3,
\qquad
W = \frac{1}{2}\,h\,\Phi_1\,\Phi_2^2+ m_s\,\Phi_2\Phi_3 
+f\,\Phi_1
\eea

\medskip\noindent
We assume that  $\Phi_{2,3}$ have a large supersymmetric mass ($m_{s}$).
$\Phi_1$ is the goldstino superfield. Its scalar
 and fermionic  components are both  massless at the tree level.   
At the quantum level  the sgoldstino  acquires a small mass.

To see this one  computes the one-loop correction to the   Kahler potential 
\medskip
\bea
K_{\rm 1-loop}=-\frac{1}{32 \pi^2}\,\tr\,\Big\{\,\cM^\dagger \cM \ln\frac{\cM^\dagger \cM}{\Lambda^2}\Big\}
\eea

\medskip\noindent
where $\cM_{ij}=M_{ij}+ \Phi_1\,N_{ij}$ is read from re-writing the above
 superpotential in the 
form $W=f\,\Phi_1+ (1/2) (M_{ij}+\Phi_1 N_{ij})\Phi_i\Phi_j$, $i,j=1,2,3$.
After integrating  out  the two massive superfields $\Phi_{2,3}$, the result is
a new $K_{\rm eff}$ and $W_{\rm eff}$ shown below
\medskip
\bea
K_{\rm eff}=\Phi_1^\dagger\Phi_1-\epsilon\,(\Phi_1^\dagger\Phi_1)^2+\cO(\epsilon^2),\qquad
W_{\rm eff}=f\,\Phi_1,
\qquad
{\rm with}
\qquad 
\epsilon=\frac{1}{12}\Big(\frac{h^2}{4\,\pi}\Big)^2\,\frac{1}{\vert m_s \vert^2}
\eea

\medskip\noindent 
This result  is valid under a simplifying 
assumption of small supersymmetry breaking: $f\,h < \vert m_s\vert^2$ (for  details see \cite{I,S,Z,B}).
As a result, the mass of the sgoldstino is small but non-zero, 
as shown by the higher dimensional D-term above  generated by quantum corrections:
$m_1^2=4\, \epsilon\,f^2$. However, for  a reliable effective theory approach, this
mass  should be of the order of supersymmetry  breaking scale $\cO(f)$.
Therefore one should have $h^2\sim \cO(4\pi)$  i.e. a nearly {\it strongly coupled} regime. 
This indicates that in general it is difficult to generate perturbatively a non-zero mass for 
sgoldstino  and this is expected to be rather light, so integrating it
out can be problematic.

As seen in the previous example of eq.(\ref{st1}) with  replacements
 $\tilde c=0$ and $c/\Lambda^2\rightarrow \epsilon$,
we find  that $ \phi_1={\psi_1\psi_1}/(-2\,f)$. As a result, on the ground state 
 the goldstino superfield satisfies again  $\Phi_1^2=0$. As before, one can check that 
$X=8/3 f\Phi_1+\cO(\epsilon)$ and $X^2=\cO(\epsilon)$, that 
vanishes in the infrared.

\subsection{Akulov-Volkov action from  nilpotent  goldstino superfield}
\label{ex3}

From the previous examples one may infer  the infrared nilpotent property of the goldstino superfield
could be  more general,  then one should be able to relate it  to the non-linear realization of  
Akulov-Volkov  action  of the goldstino \cite{AV}.  Consider then an interaction-free 
Lagrangian endowed with 
this nilpotent property
\medskip
\bea\label{ik}
\cL=\int d^4\theta\,\Phi^\dagger\Phi +
\Big\{\int d^2\theta \,f\,\Phi+{\rm h.c.}\Big\}
\qquad{\rm with}\qquad \Phi^2=0.
\eea

\medskip\noindent
The solution to this constraint is
\medskip
\bea
\Phi=\gamma\frac{\psi\psi}{ F}+\sqrt 2 \theta\psi+\theta\theta\,F\,\qquad\gamma=1/2.
\eea

\medskip\noindent
This is  used  back into eq.(\ref{ik}) to find the equation of motion for the auxiliary field
\medskip
\bea
F=-f-\gamma^2\,\frac{\opsi\opsi}{F^{\dagger 2}}\,\Box\Big[\frac{\psi\psi}{F}\Big]
\eea
with solution 
\medskip
\bea
F=-f \Big[ 1- \frac{\gamma^2}{f^4} \,\opsi\opsi\,\Box(\psi\psi) 
-\frac{\gamma^4}{f^8}\, (\psi\psi)(\opsi\opsi)\Box(\psi\psi)\Box(\opsi\opsi)\Big]
\eea

\medskip\noindent
Then the onshell Lagrangian is
\medskip
\bea\label{AV2}
\cL=-f^2 +\frac{i}{2}(\psi\sigma^\mu\,\partial_\mu\opsi-\partial_\mu\psi\sigma^\mu\opsi)
-\frac{\gamma^2}{f^2}\,\overline \psi\opsi\, \Box\,(\psi\psi)
-\frac{\gamma^4}{ f^6}\,(\psi\psi)(\opsi\opsi)\,\Box (\psi\psi) \Box(\opsi\opsi)
\eea

\medskip\noindent
For $\gamma=1/2$ a non-linear field redefinition shows \cite{Kuzenko,Kuzenko2} that eq.(\ref{AV2}) 
is equivalent to  the Akulov-Volkov action. 
The property  $\Phi^2=0$ is exact. Also,
using eq.(\ref{X}) and the equation of motion for $\Phi$, 
one has  $X=(8/3)f \Phi$, so $X^2=0$ too, and 
 there are no other superfields present.

One can also ask if  the 
Akulov-Volkov Lagrangian can be recovered by using instead a weaker constraint
 $\Phi^n\!=\!0$, $n\!>\!2$, $\Phi^2\not=0$. This has a  
solution\footnote{
$\Phi^n\!=\phi^n\!+n \sqrt 2 \theta\psi\, \phi^{n-1}+ n\phi^{n-2} [\phi F-(n-1)\psi\psi/2]$
which vanishes for any $\phi\propto \psi\psi/F$, $n>2$.} 
with $\phi=\gamma\psi\psi/F$ with $\gamma\not= 1/2$.
 However, the limit of an infinite mass of the sgoldstino 
for any $K(\Phi,\Phi^\dagger)$, $W(\Phi)$ fixes  $\gamma=1/2$ \cite{Gatto}.
So the superfield description  with $\gamma=1/2$ in eqs.(\ref{ik}), (\ref{AV2}) is 
unique\footnote{This  view is also supported by the fact that 
there seems to be no  mapping \cite{Kuzenko,Kuzenko2}  
of eq.(\ref{AV2}) with $\gamma\not=1/2$ to  the Akulov-Volkov Lagrangian.
The author thanks S. J. Tylor and  S. Kuzenko for this clarification.}.

\subsection{A counter-example to  $\Phi^2=0$}\label{2.4}

The question is how general the previous examples are 
when additional fields and interactions  are present.
Is the  goldstino superfield nilpotent when more superfields are present
 with more complicated $K$ and $W$?
The examples in Sections~\ref{ex1}, \ref{ex2} 
 seem to suggest this is indeed the case \cite{SK}. 
Consider however the following example \cite{dg2}
\medskip
\bea\label{k1}
K&=&
\Phi_1^\dagger \Phi_1 + \Phi_2^\dagger \Phi_2 
-\epsilon_1\,(\Phi_1^\dagger\Phi_1)^2
-\epsilon_2\,(\Phi_2^\dagger\Phi_2)^2
-\epsilon_3\,(\Phi_1^\dagger\Phi_1) (\Phi_2^\dagger\Phi_2)
\nonumber\\[4pt]
&-&\epsilon_4\,\big[ (\Phi_1^\dagger)^2\,\Phi_2^2+{\rm h.c.}\big]
+\cO(1/\Lambda^3)
\eea  

\medskip\noindent
and the superpotential
\bea\label{w1}
W=f\,\Phi_1, \qquad \epsilon_i=\cO(1/\Lambda^2)
\eea

\medskip\noindent
The scalar potential is
\bea
V=W_i\,(K^{-1})^i_j\,W^j
= f^2\,\big(
1+\epsilon_3\,\vert \phi_2\vert^2+4\,\epsilon_1\,\vert\phi_1\vert^2
\big)+\cO(1/\Lambda^3)
\eea

\medskip\noindent
The ground state is $\langle \phi_1\rangle =\langle \phi_2\rangle=0$ and
the scalars masses are:  $m_{\phi_1}^2= 4\epsilon_1 \,f^2$, 
 $m_{\phi_2}^2=\epsilon_3\,f^2$.
$\Phi_1$ is the goldstino superfield. 
With $\langle F_1\rangle=-f+\cO(\epsilon_i)$, $\langle F_2\rangle=\cO(\epsilon_i)$, 
after using the equations of motion to
integrate out massive $\phi_{1,2}$, one finds  that 
\medskip
\bea\label{h2.4}
\phi_1&=& 
-\frac{\psi_1\psi_1}{2\,f}-\frac{\epsilon_4}{\epsilon_1}\frac{\psi_2\psi_2}{2\,f}
+\cO(1/\Lambda)
\nonumber\\[4pt]
\phi_2&=& -\frac{\psi_1\psi_2}{f}+\cO(1/\Lambda)
\eea

\medskip\noindent
where $\phi_1$ is the sgoldstino and $\phi_2$ is a matter scalar. This gives that onshell
\medskip
\bea
(\Phi_1)^2=
\frac{\epsilon_4}{\epsilon_1}\,
\frac{\psi_2\psi_2}{(-f)}\,
\Big[\,\frac{\psi_1\psi_1}{2\,(-f)}+\sqrt 2 \,\theta\psi_1+ \theta\theta\,(-f)\Big]
+\cO(1/\Lambda)\not=0.
\eea

\medskip\noindent
The goldstino superfield does not respect the 
relation $(\Phi_1)^2=0$ in infrared\footnote{Actually this
 conclusion and similar relations to those in the text  also apply off-shell \cite{dg3,dg2}.}. 
This result does not depend on the UV scale $\Lambda$,
since in  $\epsilon_{4}/\epsilon_1$ this scale cancels out.
Could the nilpotent property still be respected? 
This would require  $\epsilon_4=0$, which could be respected
by demanding for example an additional R-symmetry.
Next, the denominator in $\phi_1$ and $(\Phi_1)^2$ is related  to 
the sgoldstino mass  $m_{\phi_1}=4\epsilon_1 f^2$ which, if very large,  
could  formally restore the  nilpotent property.
 But this would  require $\epsilon_4\ll \epsilon_1$, which
impacts on the convergence of series expansion of Kahler potential (usually
$\epsilon_{1,4}\sim 1/\Lambda^2$).
Thus, if the goldstino superfield 
interacts with other superfields, one cannot  have $(\Phi_1)^2=0$ in infrared
without further assumptions.

Note however  that a weaker condition  is instead respected   in the 
infrared, onshell and also  offshell supersymmetry
\bea\label{cubic}
(\Phi_1)^3=(\Phi_1)^2\,\Phi_2=\Phi_1\,(\Phi_2)^2=(\Phi_2)^3=0
\eea

\medskip\noindent
These relations are simple consequences of the properties of the two-dimensional Grassmann variables
(e.g. $\psi^3=0$, $\theta^\alpha\theta^\beta\theta^\gamma=0$,  etc) and are independent of the 
fields masses. Moreover,  note that these relations are symmetric in  $\Phi_1$ and $\Phi_2$, so 
this weaker nilpotent property e.g. $\Phi_i^3=0$
is not specific to the goldstino superfield.

What about  the superfield $X$, eq.(\ref{X})? From  
 the equations of motion $\overline D^2\Phi_1^\dagger=4 f +\cO(\epsilon_i)$,
$\overline D^2\Phi_2^\dagger=\cO(\epsilon_i)$, then onshell
\bea
X=4 f \Phi_1-\frac{1}{3} \Big[\Phi_1 \overline D^2\Phi_1^\dagger+\Phi_2\overline D^2\Phi_2^\dagger\Big]
+\cO(\epsilon_i)=
\frac{8}{3}\,f\,\Phi_1+\cO(\epsilon_i)
\eea
so in the infrared and onshell  $X=(8/3) f\,\Phi_1$  and $X^3=0$ but  $X^2\not=0$.

\subsection{Another counter-example to $\Phi^2=0$}
\label{2.5}

Consider another example \cite{dg3}
\medskip
\bea
K&=&
\Phi_1^\dagger \Phi_1 + \Phi_2^\dagger \Phi_2 
-\epsilon_1\,(\Phi_1^\dagger\Phi_1)^2
-\epsilon_3\,(\Phi_1^\dagger\Phi_1) (\Phi_2^\dagger\Phi_2)
+\cO(1/\Lambda^3)
\eea  
with
\bea
W=f\,\Phi_1+\frac{\lambda}{3}\,(\Phi_2)^3, \qquad \epsilon_i=\cO(1/\Lambda^2)
\eea

\medskip\noindent
$\Phi_2$ is a matter superfield that now has Yukawa couplings and
$\Phi_1$ is the goldstino superfield. Both $\phi_{1,2}$ have masses similar 
to those in previous section. Integrating them out gives 
\bea\label{op}
\Phi_1&=& 
a_{ij}\,\psi_i\psi_j+ b_{ij}\opsi_i\opsi_j+\sqrt 2 \theta\,\psi_1+\theta\theta \,F_1
\nonumber\\
\Phi_2 &=& 
c_{ij}\,\psi_i\psi_j+ d_{ij}\opsi_i\opsi_j+\sqrt 2 \theta\,\psi_2+\theta\theta \,F_2
\eea

\medskip\noindent
where a summation is understood over $i,j=1,2$ while 
the coefficients $a_{ij}, b_{ij}, c_{ij}, d_{ij}$ are not presented here (offshell they depend on
the auxiliary fields). 
Using the properties of Grassmann variables, one shows that 
this time  an even  weaker nilpotent property exists in the infrared
 (onshell and offshell) 
\medskip
\bea\label{quartic}
(\Phi_1)^4=(\Phi_1)^3\,\Phi_2=(\Phi_1)^2(\Phi_2)^2=\Phi_1\,(\Phi_2)^3=(\Phi_2)^4=0,
\qquad (\Phi_1)^2\not=0,\,\,\,\Phi_1\Phi_2\not=0.
\eea

\medskip\noindent
When $\lambda=0$, one recovers eq.(\ref{cubic}).

For the superfield $X$, by using  the equations of motion of
$\Phi_{1,2}$: $(-1/4)\overline D^2\Phi_1^\dagger + f = \cO(\epsilon_i)$ and 
$(-1/4)\overline D^2 \Phi_2^\dagger+\lambda \Phi_2^2=\cO(\epsilon_i)$, 
one finds from eq.(\ref{X})
\medskip
\bea
X= 4\, W - \frac{1}{3}\,\big[ 4 f \Phi_1+4\lambda\,\Phi_2^3\,\big]+\cO(\epsilon_i)
=\frac{8}{3}\,f\,\Phi_1+\cO(\epsilon_i)
\eea

\medskip\noindent
This verifies again $X=8/3 \,f\,\Phi_1$ in infrared and onshell, so
in this limit $X^4\!=\!0$ but $X^2\!\not=\!0$.

We see that for simple examples with  two superfields present 
 $\Phi_1^2=0$ is  not true in infrared (even  after integrating out all scalars). 
While weaker versions of the nilpotent relation are valid $\Phi_k^n=0$, $n=3,4...$, $k=1,2$
 after integrating out the scalars, this
property is not specific to the goldstino superfield anymore.  Moreover, if there are
light  scalars (that cannot be integrated  out) even this weaker version of
the nilpotent relation can fail.

\subsection{Nilpotent property  in a general case: conditions for use}

From these counter-examples it is clear that the nilpotent property 
of the goldstino superfield is not valid in general in the presence 
of more superfields and under the assumption of an initial  microscopic (UV)  Lagrangian with 
linear supermultiplets. Here we review the situation for  more 
general $K$ and $W$ and consider the following case
 \cite{dg1}\footnote{Hereafter we use superscripts  to label fields
 $\Phi^i\!=\!(\phi^i\!,\psi^i\!,F^i)$
not to be confused with  powers ($\Phi_i^\dagger$ for h.c.)}
\medskip
\bea\label{ooss}
L&=&
\int d^4\theta\,\, K(\Phi^i,\Phi_j^\dagger)+
\Big\{\int d^2 \theta\,\, W(\Phi^i)+ {\rm h.c.}
\Big\}
\nonumber\\
&=&K_i^{\,j}\,\Big[
\partial_\mu\phi^i\partial^\mu \phi_j^\dagger 
+
\frac{i}{2}\,\big(\psi^i\,\sigma^\mu\,\cD_\mu\opsi_j- \cD_\mu\psi^i\sigma^\mu\opsi_j\big)
\Big]
-W^k\,(K^{-1})^i_k\,W_i\\[4pt]
&-&\frac{1}{2}\,\Big[\big(W_{ij}-\Gamma_{ij}^m\,W_m\big)\,\psi^i\psi^j+h.c.\Big]
+\frac{1}{4}\,R_{ij}^{kl}\,\psi^i\psi^j\,\opsi_k\opsi_l,
\quad {\rm with}\quad
R_{ij}^{kl}\equiv K_{ij}^{kl}-K_{ij}^n\,\Gamma_n^{kl}
\nonumber
\eea

\medskip\noindent
in a standard notation\footnote{
We use $K_i\equiv \partial K/\partial\phi^i$,
$K^n\equiv \partial K/\partial\phi^\dagger_n$,
$K_i^n\equiv \partial^2 K/(\partial\phi^i\,
\partial \phi_n^\dagger)$,
$W_j=\partial W/\partial \phi^j$, $W^j=(W_j)^\dagger$, in which
 $W=W(\phi^i)$, $K=K(\phi^i,\phi_j^\dagger)$  are now functions of 
scalars. Note $K_{ij}^k\sim 1/\Lambda$, $K^{ij}_{km}\sim 1/\Lambda^2$, with
 $\Lambda$ the UV cutoff. 
Also  $(\Gamma_{jk}^l)^\dagger=\Gamma_l^{jk}$, $(K_{jk}^m)^\dagger=K_m^{jk}$,
$(K^{-1})_m^l=(K^{-1})^m_l$, 
$ \cD_\mu \psi^l\equiv\partial_\mu\psi^l-\Gamma^l_{jk}\,(\partial_\mu\phi^j)\,\,\psi^k$,
\, $\Gamma^l_{jk}=(K^{-1})^l_m\,K_{jk}^m$,\,
$F_m^\dagger=-(K^{-1})^i_m \,W_i +({1}/{2})\,\Gamma_m^{lj}\,\opsi_l\opsi_j$. 
In  complex geometry $R_{ij}^{kl}$ denotes
$(R_{\overline k i})_{\overline l\,j}=K_{ij\,\overline l \overline k}
-K_{ij\overline\rho}\,K^{\overline\rho n}\,K_{n\overline k\overline l}$.
}. $L$ is derived from the offshell form $\cL$\footnote{
$\cL\!\!= \!\!\!
K_i^{j} [\partial_\mu\phi^i\partial^\mu\phi_j^\dagger
\!+\!\frac{i}{2} (\psi^i\sigma^\mu\cD_\mu\opsi_j\!
 \!-\!\cD_\mu \psi^i\sigma^\mu\opsi_j)\!
+\! F^i F_j^\dagger]
\!+\!\frac{1}{4}K_{ij}^{kl}\psi^i\psi^j\opsi_k\opsi_l
\!+\! [(W_k\! -\! \frac{1}{2} K_k^{ij}\opsi_i\opsi_j) 
F^k\! -\! \frac{1}{2} W_{ij} \psi^i\psi^j\!+{\rm hc}]$}.
Denote
\medskip
\bea\label{gs1}
f_i= W_i(\langle\phi^m\rangle), 
\quad f_{ij}=W_{ij}(\langle\phi^m\rangle), 
\quad f_{ijk}=W_{ijk}(\langle\phi^m\rangle), 
 \quad f^{i}=W^i(\langle\phi^m\rangle), \,\mbox{etc.}
\eea

\medskip\noindent
In normal coordinates that we use in the following\footnote{In normal coordinates
$k^i_j=\delta_i^j$, $k^i_{jk...}=k_i^{jk....}=0$,  $R_{ij}^{kl}=k_{ij}^{kl}$
where $k^{i...}_{j....}$ are the values of $K^{i....}_{j...}$ computed on the ground state
 $\langle\phi^k\rangle$,  $\langle F^k\rangle$, $\langle\psi^k\rangle\!=\!0$
with  the field fluctuations given by
  $\delta\phi^i\!=\!\phi^i-\langle\phi^i\rangle$.}
the curvature tensor is $R_{ij}^{lm}=k_{ij}^{lm}$  where $k_{ij}^{lm}$ are the values of
$K_{ij}^{lm}$ computed on the ground state.
Further,  from   the eqs of motion for $F^i$, $\phi^i$ one has
 $ k_i^j\,\langle F_j^\dagger\rangle +f_i=0$ and 
$k_{im}^{\,j}\,\langle F^i\rangle \langle F_j^\dagger\rangle+f_{km}\langle F^m\rangle=0$
giving $\langle F_j^\dagger\rangle=-f_j$, and $f_{km}\langle F^m\rangle=0$.  SUSY breaking requires  
at least one  non-vanishing $\langle F_j\rangle$, so $\det f_{ij}=0$.    
The fermions mass matrix  $W_{ij}-\Gamma_{ij}^m\,W_m$ equals $f_{ij}$ in normal coordinates. 
The scalar mass matrix 
\bea
M_S^2=
\left[\begin{array}{lr}
\langle V^k_{\,\,\,\,l}\rangle    & \langle V_{kl}\rangle\\
\langle V^{kl}\rangle   & \langle V^{\,\,\,l}_k\rangle
\end{array}
\right]
=
\left[
\begin{array}{lr}
f^{ik}\,f_{il} - k_{il}^{jk}\,\, f^i\,f_j     &   f_{jkl}\,f^j\\
f^{jkl}\,f_j           &   f^{il}\,f_{ik} - k_{ik}^{jl}\,\,f^i\,f_j
\end{array}\right],
\label{massb}
\eea

\medskip\noindent
where $V^k_{\,\,\,l}=\partial^2 V/(\partial\phi^l\partial \phi_k^\dagger)$, 
 $V_{kl}=\partial^2 V/(\partial\phi^l\partial \phi^k)$. We assume in the following that 
\bea\label{ccon}
f_{ij}=0,\qquad  f^{ijl} f_l=0, 
 \qquad
 \mbox{and}\qquad  
 f^{ijlm}=0,
\qquad i,j=1,2. 
\eea

\noindent
which simplify the scalar mass matrix and also enable one  to integrate $\phi^i$ in terms
of massless fermions without further restriction.  
For two superfields the mass eigenvalues are
\medskip
\bea\label{pp}
m_{\tilde \phi^{1,2}}^2\!=\!\frac{1}{2}\Big[-k_{mk}^{mj}\,f^k f_j\pm\sqrt \Delta \Big],
\quad
\Delta\!=\! (k_{mk}^{mj}\,f^k f_j)^2-4\,\det( k_{mk}^{pj}\,f^k f_j)
\eea

\medskip\noindent
where det  is over the free indices and $i,j,k,m...=\!1,2$.
Consider now the transformation
\bea\label{re}
\tilde\Phi^{1}&=&\frac{1}{f}\,(f_1\,\delta
\Phi^1+f_2\,\delta \Phi^2),
\nonumber\\
\tilde\Phi^{2}&=&\frac{1}{f}\,(-f_2\,\delta \Phi^1+f_1\,\delta\Phi^2),
\qquad\,\,\, f=\sqrt{f_k f^k}
\eea

\medskip\noindent
where $\delta\Phi^j\!=(\delta\phi^j,\psi^j,F^j)$ and
$\delta\phi^j\equiv\phi^j-\langle\phi^j\rangle$.
$\tilde F^1$ is a combination of auxiliary fields
that  ``collects'' supersymmetry breaking from all directions $f^j$; this
combination of $F^j$  also dictates that of their superfields $\delta\Phi^j$ 
for spontaneous supersymmetry breaking, hence $\tilde\Phi^1$ is the  goldstino  superfield 
and $\tilde\Phi^2$ is normal to it.

Let us now discuss the decoupling of the scalars  and check
under what conditions the goldstino superfield  is nilpotent  in infrared.
We integrate the scalars which become  combinations of the massless fermions.
From  eq.(\ref{ooss}) the eq of motion of $\phi^\dagger_l$ at  zero-momentum (infrared)
becomes, after expanding it about the ground state 
\medskip
\bea\label{eqm}
k_{kj}^{il}\,\delta\phi^j\,\,f^k f_i+\frac{1}{2}\,k_{ij}^{lm}\,f_m\,\psi^i\psi^j
-\frac{1}{2}\,f^{ijl}\,\opsi_i\opsi_j+\cO(1/\Lambda^3)=0,\qquad i,j,k...=1,2.
\eea

\medskip\noindent
Taking $l=1, 2$, one solves this system for $\delta\phi^{1,2}$ to find
\medskip
\bea\label{ss22}
\delta\phi^1&=& \frac{1}{2 \det(k_{lm}^{kn}\,f_n f^m)}\,
\Big[
A_{ij}\,\,\psi^i\psi^j+ B^{ij} \,\,\opsi_i\opsi_j\Big]+\cO(1/\Lambda)
\nonumber\\
\delta\phi^2&=& \frac{1}{2\det(k_{lm}^{kn}\,f_n f^m)}\,
\Big[
 C_{ij}\,\,\psi^i\psi^j+ D^{ij} \,\,\opsi_i\opsi_j)\Big]+\cO(1/\Lambda)
\eea
with
\bea
A_{ij}&=&  \big(k_{ij}^{2p}\,k^{1 r}_{2s}-k_{ij}^{1p}\,k^{2r}_{2s}\big)\,f^r f_s f_p,
\quad
B^{ij}= -  f^{ij2}\,k_{2s}^{mr}\,f^s f_m f_r  (f_1)^{-1},
\nonumber\\
C_{ij}&=& \big(k_{ij}^{1p}\,k^{2r}_{1s}-k_{ij}^{2p}\,k^{1r}_{1s}\big)\,f^r f_s f_p,
\quad
D^{ij}=- f^{ij1}\,k_{1s}^{mr}\,f^s f_m f_r (f_2)^{-1}.
\eea

\medskip\noindent
The fields $\delta\phi^{1,2}$  are suppressed by 
the product of sgoldstino and matter
scalar masses.

Imposing the nilpotent property $(\tilde \Phi_1)^2=\cO(1/\Lambda)$, 
one finds from eqs.(\ref{re}), (\ref{ss22})
\medskip
\bea\label{dec}
f_p\,f_r f^s \big[
k_{ij}^{2p}\,(f_1\,k^{1 r}_{2 s} - f_2 \, k^{1 r}_{1 s} )-
k_{ij}^{1p}\,(f_1\, k^{2 r}_{2 s}  - f_2\,  k_{1 s}^{2 r})
\big]
+ \det (k_{n s}^{m r} f_r f^s)\,\frac{f_i f_j}{f_k f^k}
&=& 0
\nonumber\\[6pt]
k_{ls}^{mr}\,f^{ijl}\,f^s f_m f_r=0,\qquad\qquad \qquad\qquad & &
\eea

\medskip\noindent 
where  $i,j=1,2$ are fixed.
If these conditions are respected, then the goldstino superfield
satisfies  $(\tilde\Phi^1)^2=0$  in infrared, with finite scalar masses. This is true
 in the presence of superpotential interactions, with more sources of supersymmetry breaking,
after starting from  a (UV)  microscopic Lagrangian with linear superfields. Eqs.(\ref{dec})
bring constraints on the ultraviolet region controlled by the
curvature tensor $R_{ij}^{lm}=k_{ij}^{lm}$.

To illustrate further the above result, consider the simpler case of only one field 
breaking supersymmetry and take  $f_2=0$, $f_1\not=0$. 
Eqs.(\ref{ss22}) simplify
\medskip
\bea\label{st}
\delta\phi^1&=&-\frac{\psi^1\psi^1}{2\,f^1}+\frac{\det(k_{2j}^{1i})}{\det (k^{1m}_{1n})}
\,\frac{\psi^2\psi^2}{2\,f^1} 
- \frac{ \,k_{21}^{11}\,f^{ij2}}{\det (k^{1m}_{1n})}\,\frac{\opsi_i\opsi_j}{2\, \vert f_1\vert^2}
+\cO(1/\Lambda)
\nonumber\\[4pt]
\delta\phi^2&=&-\frac{\psi^1\psi^2}{f^1}
+\frac{k_{11}^{12}\,k_{22}^{11}-k_{11}^{11}\,k_{22}^{12}}{\det (k^{1m}_{1n})}
\,\frac{\psi^2\psi^2}{2\,f^1} 
+ \frac{ \,k_{11}^{11}\,f^{ij2}
}{\det (k^{1m}_{1n})}\,\frac{\opsi_i\opsi_j}{2\,\vert f_1\vert^2}
+\cO(1/\Lambda)
\eea

\medskip\noindent
which generalise the 
results in Sections \ref{2.4}, \ref{2.5}, such as eq.(\ref{h2.4}).
The terms proportional to $f^{ij2}$ are  dominant since they grow like $\Lambda^2$,
($k^{ij}_{lm}\sim 1/\Lambda^2$). The  coefficients 
of $\psi^i\psi^j$ are  independent of $\Lambda$ but 
still depend on the couplings in the microscopic Lagrangian.
Eqs.(\ref{dec}) also simplify to
\bea
\det(k^{1i}_{2j})=0,\qquad \mbox{and}\qquad f^{ij2}\,k^{11}_{12}=0.
\label{tt}
\eea
Then $(\tilde \Phi_1)^2=\cO(1/\Lambda)$
in the presence of trilinear interactions (as also seen  from
eq.(\ref{st}))\footnote{
If $\tilde\phi^1$ of (\ref{re})
is a mass eigenstate (which happens if
$k^{11}_{12}\!=\! k^{12}_{11}\!=\!0$) then  eq.(\ref{tt})  reduces to $k^{11}_{22} k_{12}^{12}\!=\!0$.}.

As a further illustration, 
consider  $W=f_1 \Phi^1+ \lambda/3!\,(\Phi_2)^3$ with a nearly massless matter scalar,
which demands $\det(k_{1m}^{1n})\approx 0$ giving
$m_{\tilde\phi_1}^2\approx -(k_{11}^{11}+k_{12}^{12}) f_1^2$, $m_{\tilde\phi_2}^2\approx 0$.
One can re-do the above calculation and integrate the sgoldstino only to find
\medskip
\bea
\delta\phi^1=-\frac{1}{ f_1\,k_{11}^{11}}\,\big[
(1/2)\,k_{mn}^{11}\,\psi^m\psi^n +f_1\delta\phi^2\,k_{12}^{11}\big]
+\cO(1/\Lambda)
\eea

\medskip\noindent
Then no power of the goldstino superfield can vanish in infrared,
unless $k^{11}_{12}=k_{22}^{11}=0$.

What about superconformal $X$? 
Consider now  that  $W=f_i\Phi^i+(1/3!) \,f_{ijk}\Phi^i\Phi^j\Phi^k$ 
and the equation of motion of $\Phi^i$: \,\,
$-1/4\,\overline D^2\Phi_i^{\dagger}+f_i+1/2\,f_{ijk}\,\Phi^j\Phi^k+\cO(1/\Lambda)=0$
where $\cO(1/\Lambda)$ accounts for higher dimensional Kahler terms. 
Then from eq.(\ref{X}) one finds, up to improvement terms:
 $X=4W -(1/3) \overline D^2 K = 8/3\,f_i\Phi^i+\cO(1/\Lambda)=(8/3)\,f \,\tilde\Phi^1$ after
 using $\overline D^2 K=\Phi^i\overline D^2 \Phi_i^\dagger+\cO(1/\Lambda)$
and eq.(\ref{re}).
This result is valid in the infrared, at scales/momenta far below $\Lambda$
and verifies that $X$ flows into goldstino superfield $\tilde\Phi^1$.
Then the same restrictions  regarding the validity of the
nilpotent constraint for $\tilde\Phi^1$   apply to $X$ too.

\subsection{Avoiding the nilpotent constraint}

The  use of the nilpotent constraint 
to define the goldstino superfield in the
presence of additional superfields can be avoided in applications.
One simply uses eq.(\ref{X}) 
which recovers in infrared the goldstino superfield, so
that offshell\footnote{Assuming the expression in eq.(\ref{X}) is indeed
valid offshell.}$^{,}$\footnote{One has $\overline D^2 \Phi_i^\dagger=
-4 (F_i^\dagger,  \,i\, \partial_\mu\sigma^\mu\opsi_i, - \Box\phi_i^\dagger)$ 
for $\Phi^i=(\phi^i,\psi^i,F^i)$
and that  $K=\Phi_i^\dagger\Phi^i+\cO(1/\Lambda)$.}
\medskip
\bea\label{ll1}
\tilde \Phi^1= \frac{3}{8 f}\Big[
 4 W(\Phi^i)+\frac{4}{3}\,\Phi^i\, \big( F_i^\dagger+ i\sqrt 2\, \theta \,\partial_\mu\sigma^\mu\opsi_i 
-\theta\theta\, \Box\phi_i^\dagger\,\big)+\cO(1/\Lambda)\Big]
\eea

\medskip\noindent
with $f=\sqrt{f_i\,f^i}$ and its onshell form 
\medskip
\bea\label{ll2}
\tilde \Phi^1=\frac{3}{8 \,f}
\Big[
4 W(\Phi^i)-\frac{4}{3} \Phi^i\,\frac{\partial W}{\partial\Phi^i}+\cO(1/\Lambda)
\Big]
\eea

\medskip\noindent
from which the expression of sgoldstino is obvious.
This form  does not integrate out the
other scalar fields\footnote{When this is possible, the sgoldstino 
becomes a combination of Weyl bilinears, as seen earlier.}. 
Eqs.(\ref{ll1}), (\ref{ll2})  do not depend on  $K$ in  leading order and can be used
in applications that need the expression of the goldstino superfield and its sgoldstino
in infrared.

\section{Conclusions}

Superfield constraints are often used  to project out
some degrees of freedom of the microscopic Lagrangian and provide  a non-linear 
realization of supersymmetry.  The Akulov-Volkov is  the 
celebrated  example, described by a free action with
 $L=[\Phi^\dagger\Phi]_D+[f\Phi_F+{\rm h.c.}] $,  with a constraint
for  goldstino superfield 
 $\Phi^2\!=\!0$ and  solution $\Phi\!=\!\psi\psi/(2F)\!+\!\sqrt 2 \theta\psi\!+\!\theta\theta\,F$. 
The constraint projects out the sgoldstino which becomes a bilinear of goldstini. 
$L$ above  was initially derived perturbatively from  a  general, microscopic
Lagrangian for $\Phi$ in which one decouples the sgoldstino (by taking its mass to infinity),
leading to the above solution for $\Phi$ and thus to $L$.

Inspired by this result, the nilpotent property  $\Phi^2=0$ is sometimes  used in the literature
to define the goldstino superfield even  {\it in the presence} 
of additional superfields $\Phi^i=(\phi^i,\psi^i,F^i)$.
This procedure can lead to incorrect results.
We first reviewed  counter-examples to $\Phi^2=0$ in the infrared when starting
with  microscopic Lagrangians with two linear superfields present.
We identified the goldstino superfield and checked the nilpotent 
property after integrating  out {\it all scalars} of the theory;
then the sgoldstino becomes a combination of Weyl bilinears
$\phi=a_{ij} \psi^i\psi^j+b^{kl}\opsi_k\opsi_l$
and the nilpotent property is not respected.
In some cases a weaker version of this property is valid such
as $\Phi^3=0$ or $\Phi^4=0$ with  $\Phi^2\not=0$. 
The reason is simple: for large enough powers $n$, $\Phi^n$ vanishes 
in the infrared, once its scalar component is of the type shown above.
Moreover,  this weaker property applies to both superfields i.e. it is not
specific to the goldstino superfield.

We then reviewed more general cases. 
Assuming linear supermultiplets and  a microscopic
Lagrangian with  general $K(\Phi^i,\Phi_i^\dagger)$ and  $W(\Phi^i)$ and more sources 
of supersymmetry breaking,  we found the conditions for the
Kahler curvature tensor and superpotential couplings under 
which one can still have  $\Phi^2=0$ in the infrared, after integrating out the scalars. 
These conditions are actually very restrictive for the model and  for the ultraviolet  
region (controlled by the Kahler curvature tensor). If massless scalars exist
(coupled to sgoldstino) even the weaker versions of the nilpotent property can be
invalidated.

The use of the nilpotent condition  of the goldstino superfield can be avoided.
Since  the  superconformal symmetry breaking chiral superfield $X$ 
goes in infrared to the goldstino superfield, one can just use its known
expression (determined by $K$, $W$ and only by $W$ up to $\cO(1/\Lambda)$)
 to obtain the (infrared) sgoldstino  in terms of the  other scalars. This expression
can then be used in applications.  The scalars can eventually
be integrated out, so the sgoldstino becomes a combination
of Weyl fermions bilinears, leading  to $\Phi^2\not=0$ 
unless the aforementioned conditions are respected.
 This is the picture if one assumes the initial existence of a 
perturbative UV description of the  theory with linear supermultiplets.

One can also proceed in the opposite way:  simply use the 
nilpotent property  $\Phi^2~=~0$ as a definition for the goldstino multiplet even in 
the presence  of additional superfields. Doing so is   popular e.g.  
in ``nilpotent inflation'' models in which goldstino superfield is present,  since 
in this case the sgoldstino becomes a bilinear  of goldstini and 
one does not have to stabilize this field direction,  simplifying  
the calculation. A good question to ask is then whether the nilpotent relation which  
initially was only an {\it infrared} property (on the ground state) 
can be extended  (to non-zero momenta, at the  scale of inflation, etc).
Questions can also  be asked on the nature of supersymmetry breaking in this case
 and  on the  microscopic (UV) description of such models in terms of linear supermultiplets.
 In fact such UV completion   might not exist perturbatively
(unless the restrictive  conditions mentioned earlier are respected).

\vspace{1cm}
\noindent
{\bf Acknowledgements: }
The author thanks S. J. Tylor and S. Kuzenko for a discussion.
This work  was supported  by the National  Authority for Scientific 
Research (CNCS-UEFISCDI) under project number PN-II-ID-PCE-2011-3-0607.

\end{document}